\documentclass[fleqn,twoside]{article}
\usepackage{espcrc2}

\usepackage{graphicx}
\usepackage[figuresright]{rotating}

\newcommand{\AmS}{{\protect\the\textfont2
  A\kern-.1667em\lower.5ex\hbox{M}\kern-.125emS}}

\title{CP Violation in $B$ Meson Decays\thanks{Talk presented 
at the Sixth International Conference on Hyperons, 
Charm and Beauty Hadrons, IIT, Chicago, June 27--July 3 2004.}}

\author{Michael Gronau\address{Physics Department, Technion -- Israel 
Institute of Technology, \\ 
        32000 Haifa, Isreal}}%
       
\begin{document}

\def \app{D_{\pi \pi}}
\def \b{{\cal B}}
\def \bbpp{\overline{{\cal B}}_{\pi \pi}}
\def \bea{\begin{eqnarray}}
\def \beq{\begin{equation}}
\def \bg{\bar \Gamma}
\def \bl{\bar \lambda}
\def \bo{B^0}
\def \ko{K^0}
\def \ob{\overline{B}^0}
\def \lesssim{\stackrel{<}{\sim}}
\def \largesim{\stackrel{>}{\sim}}
\def \bpb{\stackrel{(-)}{B^0}}
\def \cn{Collaboration}
\def \cpp{C_{\pi \pi}}
\def \eea{\end{eqnarray}}
\def \eeq{\end{equation}}
\def \ite{{\it et al.}}
\def \kpb{\stackrel{(-)}{K^0}}
\def \lpp{\lambda_{\pi \pi}}
\def \ob{\overline{B}^0}
\def \ok{\overline{K}^0}
\def \rpp{R_{\pi \pi}}
\def \rt{r_{\tau}}
\def \s{\sqrt{2}}
\def \half{\frac{1}{2}}
\def \3half{\frac{3}{2}}
\def \spp{S_{\pi \pi}}

\begin{abstract}
Recent CP asymmetry measurements in 
tree-dominated processes, $B^0\to \pi^+\pi^-, \rho^+\rho^-, \rho^{\pm}\pi^{\mp}, 
B^+\to DK^+$, and in penguin-dominated decays, $B \to \pi^0 K_S, \eta' K_S, \phi K_S$, 
are interpreted in the framework of the Kobayashi-Maskawa (KM) mechanism of CP 
violation. The KM phase emerges as the dominant source of CP violation in 
tree-dominated decays, which are beginning to constrain the unitarity triangle
beyond other constraints. Improving precision of CP asymmetry measurements 
in penguin-dominated decays may 
indicate the need for new physics. 

\vspace{1pc}
\end{abstract}

\maketitle

\section{INTRODUCTION}

Measurements of time-dependent CP asymmetries in $b \to c \bar cs$
decays including $B^0 \to J/\psi K_S$, carried out in the past few years at the 
two $e^+e^-$ $B$ factories at SLAC and KEK~\cite{psiKs}, are interpreted 
in the Standard Model as $\sin 2\beta\sin\Delta mt$, where
$\beta \equiv \arg(-V_{tb}V^*_{td}V_{cd}V_{cb}^*)$. These measurements
were proposed in~\cite{BS} to provide a first test of the Kobayashi-Maskawa
mechanism of CP violation~\cite{KM} in the $B$ meson system. This test is theoretically clean 
because a single weak phase dominates $B\to J/\psi K_S$ within a fraction of a 
percent~\cite{MG,LonPec}. These measurements have not only passed successfully 
the Standard Model test; they also improve our knowledge of the
Cabibbo-Kobayashi-Maskawa (CKM) matrix parametrized in terms of the unitarity 
triangle. (See Fig.~1~\cite{CKMfitter}.) 
A recent time and angular analysis of $B^0\to J/\psi K^{*0}$~\cite{psiK*}  seems to 
resolve the plotted ambiguity, $\beta\to \pi/2 -\beta$. 

CP asymmetries in $B$ 
decays are often related to the three angles of the unitarity triangle,
for which currently allowed ranges are, at 95$\%$ confidence level 
(CL)~\cite{CKMfitter}:
\beq\label{albega}
78^\circ \le \alpha \le 122^\circ,\,21^\circ \le \beta \le 27^\circ,
38^\circ \le \gamma \le 80^\circ
\eeq
\begin{figure}[h]
\centerline{\includegraphics[height=4.2cm,width=6.0cm]{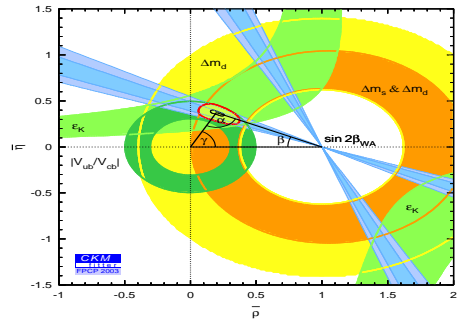}}
\caption{CKM constraints~\cite{CKMfitter}.}
\label{fig:CKM}
\end{figure}
The essence of measuring CP asymmetries in a variety of $B$ and $B_s$ 
decays is first to 
see whether the measured asymmetries are consistent with (\ref{albega}).
Then one would like to restrict these ranges further, hoping
to eventually observe deviations from Standard Model expectations. 
A class of processes, susceptible to a possibe early observation of new physics, 
mimicking loop effects, consists of $b \to s$ penguin-dominated $B^0$ decays into 
CP-eigenstates~\cite{NP}, where CP asymmetries may deviate from $\sin 2\beta$.

While the phase $2\beta$ characterizes CP violation in the interference between $\bo-\ob$ 
mixing and a $b \to c \bar cs$  decay amplitude or a $b \to s$ penguin-dominated amplitude, 
the phase $\gamma \equiv  \arg(-V_{ub}V^*_{ud}V_{cd}V_{cb}^*)$
is responsible for direct CP violation. The phase $\alpha \equiv \pi - \beta - \gamma$ 
occurs when $\bo-\ob$ mixing interferes with $b \to u \bar ud$, where an additional 
$b\to d$ penguin amplitude implies also direct CP violation.
Since a direct CP asymmetry involves a ratio of two hadronic amplitudes and their relative strong phase, 
both of which are not reliably calculable, one faces hadronic 
uncertainties whenever one is relating CP asymmetries to $\gamma$ or $\alpha$. 

A model-independent way of resolving or at least reducing these uncertainties is by 
flavor symmetries, isospin or 
broken SU(3), relating certain processes to others.  
SU(3) breaking corrections  are
introduced using QCD factorization results proven in a framework 
of a Soft Collinear Effective Theory~\cite{BBNS}. 

An impressive progress is being reported at this conference in measurements of
CP asymmetries~\cite{tosi,schwartz,schrenk}. The purpose of my talk is to study
theoretical implications of these measurements. In Section 2 I discuss tree 
dominated decays, including $B^0(t)\to \pi^+\pi^-, \rho^+\rho^-, \rho^{\pm}\pi^{\mp}$. 
Section 3 studies pure tree decays $B^\pm \to D K^{\pm}$, while Section 4 focuses 
on penguin-dominated processes, $B^0(t)\to \pi^0 K_S, \eta' K_S, \phi K_S$. 
Section 5 concludes.

\section{TREE DOMINATING DECAYS}

\subsection{$B^0 \to \pi^+\pi^-$ and $B\to\rho^+\rho^-$}

The amplitude for $B^0\to\pi^+\pi^-$ contains two terms~\cite{MG,LonPec}, 
conventionally denoted ``tree" and ``penguin" amplitudes, $T$ and $P$, 
involving a weak phase $\gamma$ and a strong phase $\delta$: 
\beq
A(\bo \to \pi^+ \pi^-) =  |T| e^{i \gamma} + |P| e^{i \delta}.
\eeq
The time-dependent decay rates, for an initial $B^0$ or a $\ob$, are
given by~\cite{MG}
\bea\label{SC}
&&\Gamma(B^0(t)/\ob(t)\to\pi^+\pi^-) \propto
\nonumber\\
&&e^{-\Gamma t}
\left [ 1\pm \cpp\cos\Delta(m t) \mp \spp\sin(\Delta m t)\right ]
\eea 
where
\bea
\spp & =  & \frac{2 {\rm Im}(\lpp)}{1 + |\lpp| ^2} =
\sqrt{1- C^2_{\pi\pi}}\sin 2\alpha_{\rm eff},\\
\cpp & = &  \frac{1 - |\lpp|^2}{1 + |\lpp|^2},\\
\lpp & \equiv &  e^{-2i \beta} \frac{A(\ob \to \pi^+ \pi^-)}
{A(B^0 \to \pi^+ \pi^-)}.
\eea
Recent measurements by the BaBar~\cite{Bapipi}  and  Belle~\cite{Bepipi}
collaborations, which Belle used to announce evidence for CP violation, are
$$
\begin{array}{cccc}  & {\rm BaBar} & {\rm Belle} \\
\cpp & -0.19\pm 0.19 \pm 0.05 & -0.58\pm 0.15 \pm 0.07, \\
\spp & -0.40\pm 0.22 \pm 0.03 & -1.00 \pm 0.21 \pm 0.07.
 \end{array}
 $$
These values imply averages,
 \beq
 \cpp = -0.46 \pm 0.13,~~\spp = -0.74 \pm 0.16,
 \eeq
which also determine a range for $\sin 2\alpha_{\rm eff}$.

\medskip\noindent
{\it 2.1.1~Isospin symmetry in $B\to\pi\pi$}

Since the two measurables, $\spp$ and $\cpp$, depend on $r \equiv |P/T|, 
\delta, \gamma$ (and $\beta$ which may be assumed to be given), one 
needs further 
information to study the weak phase. This information is provided by isospin 
symmetry~\cite{GL}, which distinguishes between tree and penguin
amplitudes, also relating a tiny electroweak penguin term to the tree  
amplitude~\cite{EWP}. One forms an isospin triangle for $B$ decays
and a similar one for $\bar B$ decays,
\beq\label{pipi}
A(\pi^+\pi^-) + \s A(\pi^0\pi^0) = \s A(\pi^+\pi^0).
\eeq
A mismatch angle
between the two triangles determines $2(\alpha_{\rm eff} - \alpha)$,
which then fixes $\alpha$ from $\sin 2\alpha_{\rm eff}$ up to a
discrete ambiguity.

As long as separate $B^0$ and $\ob$ decays into $\pi^0\pi^0$ have not yet
been measured, one may try to use the three measured charge averaged decay 
rates for bounds on $\alpha_{\rm eff} - \alpha$~\cite{GQ}. Current branching 
ratios of the three processes in (\ref{pipi}) are, in units of $10^{-6}$~\cite{HFAG},
$4.6\pm 0.4, 1.9 \pm 0.5$ and $5.2 \pm 0.8$, respectively. These values 
imply $|\alpha_{\rm eff} - \alpha| < 49^\circ$ at 90$\%$ CL,
which is not very useful because of the sizable branching ratio into $\pi^0\pi^0$.

\medskip\noindent
{\it 2.1.2~Isospin symmetry in $B\to\rho\rho$}

The two $\rho$ mesons in $B\to \rho\rho$ are described by three possible 
polarization states, longitudinal or transverse to the momentum direction,
parallel or orthogonal to each other in the case of transversely polarized states. 
Angular distributions of the outgoing pions measured by
BaBar~\cite{Balong,BarhoSC} show that the $\rho$ mesons are dominantly longitudinally
polarized, $\Gamma_L/\Gamma = 0.99 \pm 0.03^{+0.04}_{-0.03}$, describing CP-even 
states. That is, the case of $B^0\to\rho^+\rho^-$ is
almost identical to the case of $B^0\to \pi^+\pi^-$, albeit a possible small correction from 
a CP-odd state, and a slight violation of the $B\to\rho\rho$ isospin relation when the two 
$\rho$ mesons are observed at different invariant masses~\cite{FLNQ}.

Time-dependent asymmetries in $B^0\to\rho^+\rho^-$, analogous to $\cpp$ and $\spp$ 
in (\ref{SC}), were measured by BaBar for longitudinally polarized $\rho$'s~\cite{BarhoSC},
\bea\label{CSrhorho}
C_{\rho\rho} & = & -0.17 \pm 0.27\pm 0.14,\nonumber\\
S_{\rho\rho} & = & -0.42 \pm 0.42 \pm 0.14.
\eea
This implies two possible central values $\alpha_{\rm eff} = 103^\circ, 167^\circ$
with large experimental errors. 
A small branching ratio of $B\to\rho^0\rho^0$,  $\b(\rho^0\rho^0) < 2.1\times 
10^{-6}$~\cite{r0r0}, much smaller than $\b(\rho^+\rho^-) = 
(25 \pm 9)\times 10^{-6}$~\cite{BarhoSC}  
and $\b(\rho^+\rho^0) = (26 \pm 6)\times 10^{-6}$~\cite{r0r0,Beller+r0}, 
implies a $90\%$ CL 
upper bound $|\alpha_{\rm eff} - \alpha| < 17^\circ$~\cite{GQ}. 
This bound and (\ref{CSrhorho}) exclude the range 
$19^\circ \le \alpha \le 71^\circ$ at 90$\%$ CL~\cite{BarhoSC} consistent with (\ref{albega}).

\medskip\noindent
{\it 2.1.3~Broken SU(3) symmetry for $B\to\pi^+\pi^-$}

As long as separate $B^0$ and $\ob$ decays into $\pi^0\pi^0$ have not yet
been measured, one may replace isospin symmetry by the less precise flavor 
SU(3) symmetry relating $B\to \pi\pi$ to $B\to K\pi$~\cite{SU3,GHLR}. This 
imposes constraints on the ratio of penguin-to-tree amplitudes in $B\to\pi\pi$,
$r \equiv |P/T|$. To improve the quality of the analysis, one 
introduces SU(3) breaking in tree amplitudes in terms of a ratio of $K$
and $\pi$ decay constants, as given by factorization~\cite{BBNS}, neglecting a 
very small variation of the $B$ to $\pi$ form factor when $q^2$ varies from 
$m_\pi^2$ to $m_K^2$. This assumption and the neglect of small annihilation 
amplitudes may be checked experimentally. Here we follow briefly
a study presented recently in~\cite{GRpipi}, which contains further  
references.

Writing [$\bl \equiv \lambda/(1 - \lambda^2/2) = 0.230$]
\bea\label{AKpi+}
A(B^+ \to K^0\pi^+) & = & -\bl^{-1}Pe^{i\delta},
\\
\label{AKpi-}
A(B^0 \to K^+\pi^-) & = & - \frac{f_K}{f_\pi}\bl\,Te^{i\gamma} + 
\bl^{-1}Pe^{i\delta},\nonumber
\eea
one defines two ratios of charged averaged rates,
\bea\label{R}
{\cal R}_+&\equiv & \frac{\bl^2\,\bar\Gamma(B^+\to K^0\pi^+)}
{\bar\Gamma(B^0\to \pi^+\pi^-)}
= \frac{r^2}{\rpp},
\\
{\cal R}_0&\equiv & \frac{\bl^2\,\bar\Gamma(B^0 \to K^+\pi^-)}
{\bar\Gamma(B^0\to \pi^+\pi^-)}
= \frac{r^2 + 2r\bl'^2z
+ \bl'^4}{\rpp}\nonumber
\eea
where $\rpp \equiv 1 - 2rz + r^2$ and
\beq
\bl'  \equiv \sqrt{\frac{f_K}{f_\pi}}\,\bl,~~z \equiv \cos\delta\cos(\beta + \alpha).
\eeq

Current experimental values~\cite{HFAG},
\beq\label{R+,0}
{\cal R}_+ = 0.235 \pm 0.026,~~{\cal R}_0 = 0.209 \pm 0.020,
\eeq
imply a large penguin amplitude 
\beq
0.51 \le r \le 0.85~~({\rm assuming}~|\delta| < \pi/2).
\eeq 
This range is consistent with a recent global SU(3) fit
to charmless $B$ decays into two pseudoscalar mesons, 
$B\to PP$~\cite{BPP,suprun}.
However, it is somewhat in conflict with most QCD-based 
calculations which typically obtain smaller values for $r$~\cite{QCD}.

A study of $\alpha$ using the CP asymmetries $\cpp$ and $\spp$ 
proceeds as follows. One expresses the two asymmetries in terms of $r$, 
$\delta$ and $\alpha$,
\bea\label{C}
\cpp & = & \frac{2r\sin\delta\sin(\beta +\alpha)}{\rpp},
\\
\label{S}
\spp & = & \frac{\sin 2\alpha + 2r\cos\delta\sin(\beta-\alpha) - 
r^2\sin 2\beta}{\rpp}.\nonumber
\eea
\begin{figure}[h]
\centerline{\includegraphics[height=6.3cm,width=7.0cm]{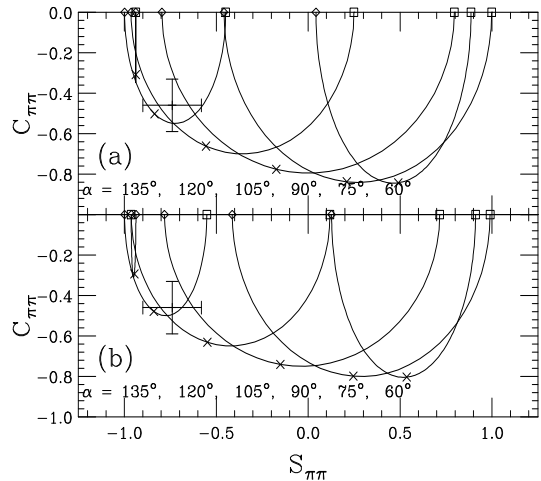}}
\caption{Curves of $C_{\pi \pi}$ vs.\ $S_{\pi \pi}$.}
\end{figure}
We now impose in addition the SU(3)
constraints (\ref{R}) and (\ref{R+,0}). 
The resulting two plots of $\cpp$ vs. 
$\spp$~\cite{GRpipi}, shown in Fig.~2(a) using ${\cal R}_+$ and 
in Fig.~2(b) using ${\cal R}_0$,
are rather similar, supporting the SU(3) 
assumption in these processes. 
The plotted experimental average (7) implies an allowed range, $\alpha = 
(104\pm 18)^\circ$, which overlaps nicely with the range of $\alpha$ in (\ref{albega}), 
favoring large values in this range. 

SU(3) breaking corrections in penguin amplitudes 
modify the allowed range of $\alpha$. The bounds become 
stronger (weaker) if SU(3) breaking enhances (suppresses) the penguin 
amplitude in $B\to K\pi$ relative to $B\to\pi\pi$. A $22\%$ SU(3) correction, 
implied for instance by a factor $f_K/f_\pi$ (or by its inverse), 
modifies the bounds by $8^\circ$. 

\subsection{Broken SU(3) symmetry for $B\to\rho^{\pm}\pi^{\mp}$}

Very recently broken flavor SU(3) has also been applied to study $\alpha$ in
$B^0(t)\to\rho^{\pm}\pi^{\mp}$~\cite{GZ}. The number of hadronic
parameters and the number of measurable quantities are twice as 
large as in $B^0\to\pi^+\pi^-$ which complicates somewhat the analysis.
Here we discuss only a bound on $\alpha$, while a complete determination
of $\alpha$ will be presented by J. Zupan at this conference~\cite{zupan}. 

Time-dependent decay rates for initially $B^0$ decaying into 
$\rho^\pm\pi^\mp$ are given by~\cite{MGPLB},
\bea\label{Gammat}
\Gamma(B^0(t) \to \rho^\pm\pi^\mp) \propto  1 + (C \pm \Delta C)\cos\Delta mt
\nonumber\\ 
 - (S \pm \Delta S)\sin\Delta mt,
\eea
while for initially $\ob$ decays the $\cos\Delta mt$ and $\sin\Delta mt$ terms have 
opposite signs.
As in $B\to\pi^+\pi^-$, one defines a measurable phase, $\alpha_{\rm eff}$, which 
equals $\alpha$ in the limit of vanishing penguin amplitudes, 
\bea
4\alpha_{\rm eff} \equiv  \arcsin\left [(S + \Delta S)/\sqrt{1- (C + \Delta C)^2}\right ]
\nonumber\\
 + \arcsin\left [(S - \Delta S)/\sqrt{1- (C - \Delta C)^2}\right ].
 \label{alphaeff}
\eea

The difference $|\alpha_{\rm eff} - \alpha|$, which is governed by penguin 
contributions, can be shown to be bounded by ratios of strangeness changing 
decay rates of $B\to K^*\pi$ and $B\to K\rho$ (dominated by penguin amplitudes)  
and decay rates of $B\to \rho^+\pi^-$ and $B\to\rho^-\pi^+$ (dominated by tree amplitudes).
In particular, using the following two ratios of rates,
\bea
{\cal R}^+_+ &  \equiv  \bl^2\Gamma(B^+\to K^{*0}\pi^+)/
\Gamma(B^0 \to \rho^+\pi^-),\nonumber\\
{\cal R}^0_- & \equiv \bl^2\Gamma(B^0\to \rho^- K^+)
\Gamma(B^0\to \rho^-\pi^+),
\eea
one may show that 
\beq
2|\alpha_{\rm eff} - \alpha| \le \arcsin\sqrt{{\cal R}^+_+}
+ \arcsin\sqrt{{\cal R}^0_-}.
\eeq
Using current values~\cite{HFAG}
\beq\label{expR}
{\cal R}^+_+ = 0.032 \pm 0.007,
~{\cal R}^0_- = 0.047 \pm 0.015,
\eeq
one obtains an upper bound at 90$\%$ CL,
\beq\label{Delta-alpha}
|\alpha_{\rm eff} - \alpha| \le 12^\circ.
\eeq
This bound, which assumes exact SU(3) for penguin amplitudes,
becomes stronger (weaker) if $\Delta S=1$ penguin amplitudes are 
enhanced (suppressed) relative to $\Delta S=0$ penguin amplitudes. 
In any event, one expects this bound to change by no more than 30$\%$,
implying $|\alpha_{\rm eff} - \alpha| \le 15^\circ$ in the presence of 
SU(3) breaking corrections.

Both BaBar~\cite{CKMfitter,BaBarrhopi} and Belle~\cite{schwartz} measured 
time-dependence in $B^0\to\rho^{\pm}\pi^{\mp}$:
\beq\label{CSDelta}
C=\left\{ \begin{array}{c}0.35 \pm 0.14 \cr
0.25\pm 0.17 \end{array}\right.
\Delta C=\left\{ \begin{array}{c}0.20 \pm 0.14 \cr
0.38\pm 0.18\end{array} \right.
\eeq
\beq
S=\left\{ \begin{array}{c}-0.13 \pm 0.18 \cr
-0.28\pm 0.24\end{array} \right.
\Delta S=\left\{ \begin{array}{c}0.33 \pm 0.18 \cr
-0.30\pm 0.26\end{array} \right.
\eeq
where the first values are BaBar's and the second are Belle's. 
These values imply single solutions,
\beq\label{aleff}
\alpha_{\rm eff} = \left\{ \begin{array}{c}(93  \pm 7)^\circ ~~~{\rm BaBar}\cr
(102\pm 11)^\circ ~~~{\rm Belle},\end{array} \right.
\eeq
when making a mild and experimentally testable assumption~\cite{GZ}
that the two arcsin's in (\ref{alphaeff}) differ by much less than $180^\circ$. 
Combining (\ref{aleff}) with the above upper bound on $|\alpha_{\rm eff} - 
\alpha|$ one obtains
\beq
\alpha=\left\{ \begin{array}{c}(93  \pm 7 \pm 15)^\circ = (93 \pm 17)^\circ~{\rm BaBar}\cr
(102\pm 11\pm 15)^\circ = (102 \pm 19)^\circ{\rm Belle}\end{array} \right.~
\eeq
where experimental and theoretical errors are added in quadrature.
These values are in good agreement with the range of 
$\alpha$ in (\ref{albega}) and largely overlap with this range. 

One may show that the relative contributions of penguin amplitudes in 
$B^0\to\rho^{\pm}\pi^{\mp}$ are much smaller than in 
$B^0\to\pi^+\pi^-$, involving ratios of penguin and tree amplitudes,
$r_{\pm} \sim 0.2$~\cite{BPP,QCD,GZ,CGLRS}. 
Consequently, effects of SU(3) breaking in penguin amplitude lead to an
intrinsic uncertainty of only a few degrees in the determination of $\alpha$ 
in $B^0(t)\to\rho^{\pm}\pi^{\mp}$~\cite{GZ}. 

\section{PURE TREE DECAYS $B^{\pm}\to DK^{\pm}$}

The process $B^+\to DK^+$ and its charge conjugate provide a way of 
determining $\gamma$ in a manner which is pure in principle, avoiding 
uncertainties in penguin amplitudes. Here we wish to study a 
scheme proposed in~\cite{GW}, which uses both $D^0$ 
CP-eigenstates and $D^0$ flavor states. An extensive list of variants of this 
method is given in~\cite{variants}. In all variants one makes 
use of an interference between tree amplitudes in decays of the type 
$B^\pm\to DK^\pm$, from $\bar b \to \bar c u\bar s$ and $\bar b \to \bar 
u c\bar s$, for which the weak phase difference is $\gamma$. We choose to 
also discuss briefly one variant~\cite{GGSZ}, in which the Dalitz plot 
of $D^0\to K_S\pi^+\pi^-$ is being analyzed in $B^{\pm}\to DK^{\pm}$ .

We denote  by $r$ the magnitude of the ratio of two amplitudes, $A(B^+\to D^0 K^+)$ 
from $\bar b \to \bar u c\bar s$ and $A(B^+\to \bar D^0 K^+)$ from $\bar b \to \bar 
c u\bar s$, and we denote by $\delta$ the strong phase of this ratio.  One then obtains 
the following expressions for two ratios of rates, for even and odd $D^0$-CP states, 
and for two corresponding CP asymmetries:
\bea
R_{\pm} & = & \frac{\Gamma(D^0_{{\rm CP}\pm} K^-) + 
\Gamma(D^0_{{\rm CP}\pm} K^+)}{\Gamma(D^0 K^-)}
\nonumber\\
& = & 1 + r^2 \pm 2r\cos\delta\cos\gamma,
\\
A_{\pm} & = & \frac{\Gamma(D^0_{{\rm CP}\pm} K^-) - 
\Gamma(D^0_{{\rm CP}\pm} K^+)}{\Gamma(D^0_{{\rm CP}\pm} K^-) 
+ \Gamma(D^0_{{\rm CP}\pm} K^+)} 
\nonumber\\
& = & \pm 2r \sin\delta \sin\gamma/R_{\pm}.
\eea
In principle, these three independent observables determine $r,~\delta$
and $\gamma$. However, this is difficult in practice since one must be 
sensitive to an $r^2$ term, where the current 90$\%$ CL upper limit 
on $r$ is $r<0.22$~\cite{DKBaBar}. A crude estimate is 
$r\sim 0.2$~\cite{BDK}, since this ratio involves a CKM factor,
 $|V^*_{ub}V_{cs}/V^*_{cb}V_{us}| = 0.4-0.5$, and probably a 
 comparable color-suppression factor. 

Taking averages of Belle and BaBar measurements when 
both are available, one finds~\cite{dataDK} 
\bea
R_+&=&1.09 \pm 0.16~({\rm Belle~\&~BaBar}),
\nonumber\\
R_-&=&1.30 \pm 0.25~({\rm Belle}),
\nonumber\\
A_+&=&0.07 \pm 0.13~({\rm Belle~\&~BaBar}),
\nonumber\\
A_-&=&-0.19 \pm 0.18~({\rm Belle}).
\eea
This implies
\beq
r = 0.44^{+0.14}_{-0.22},~~|A_{\pm}|_{\rm ave} = 0.11 \pm 0.11.
\eeq

In order to obtain constraints on $\gamma$ we eliminate $\delta$, 
plotting in Fig.~3 $R_\pm$ versus $\gamma$ for allowed $A_\pm$. 
We are using $1\sigma$ bounds on $r,~R_\pm$ and $|A_{\pm}|_{\rm ave}$.
The ratios $R_+$ and $R_-$ are described 
by the lower and upper branches, respectively, corresponding to
$\cos\delta\cos\gamma < 0$. This implies a very strong 
$1\sigma$ lower bound, $\gamma > 72^\circ$, and requires $\cos\delta < 0$ for
allowed values of $\gamma$. More precise measurements of 
$R_\pm$ are needed for constraints beyond $1\sigma$.
\begin{figure}
\centerline{\includegraphics[height=6.0cm,width=7.0cm]{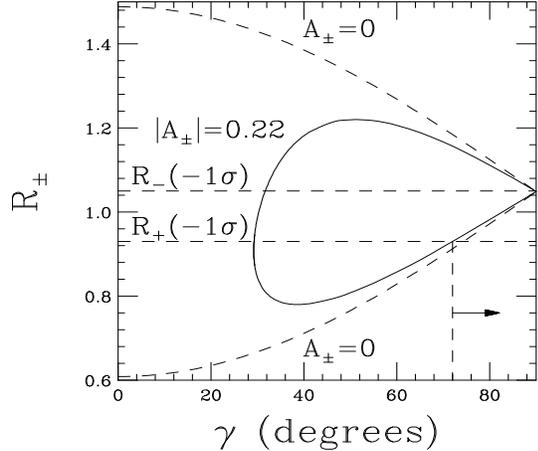}}
\caption{$R_{\pm}$ as functions of $\gamma$ for $r=0.22$ and 
$|A_{\pm}|=0.22$ (solid curve) or $A_{\pm}=0$ (dashed curve). Horizontal 
dashed lines denote $1\sigma$ experimental lower limits of $R_-$ and 
$R_+$.}
\end{figure}

A recent study by Belle of $B^\pm\to D^{(*)}K^\pm$~\cite{schrenk,BelleDalitz}, 
analyzed the Dalitz plot for 
$D\to K_S\pi^+\pi^-$ in terms of a sum of a non-resonant term and 
a set of resonances given by Breit-Wigner forms. This study, involving some
model-dependence, confirmed $\cos\delta < 0$ and  
obtained $2\sigma$ bounds, $26^\circ < \gamma < 126^\circ$,
considerably wider than  and including the range of $\gamma$ in (\ref{albega}).

\section{PENGUIN DOMINATED DECAYS}

In a class of penguin-dominated $B^0$ decays into CP-eigenstates, including
the final states  $f=\pi^0 K_S,~\eta' K_S,~\phi K_S$ and 
$(K^+K^-)_{({\rm even}~\ell)}K_S$, decay amplitudes contain two terms: a 
penguin amplitude, $p'_f$, involving a dominant 
CKM factor $V^*_{cb}V_{cs}$,  and a color-suppressed tree
amplitude, $c'_f$, with a smaller CKM factor $V^*_{ub}V_{us}$. The first
amplitude by itself would imply a CP asymmetry of magnitude 
$\sin 2 \beta\sin\Delta mt$. The second amplitude modifies 
the coefficient of this term, and introduces a $\cos\Delta mt$ term in the 
asymmetry~\cite{MG}. The coefficients of $\sin\Delta mt$ and $\cos\Delta mt$ for 
a final state $f$ are denoted by $S_f$ and $-C_f$, respectively, as given
in Eq.~(\ref{SC}) for $B\to\pi^+\pi^-$. The observables, $\Delta S_f \equiv S_f \pm
\sin 2\beta$ (where the sign depends on the final state CP)
and $C_f$, increase with $|c'_f/p'_f|$, but
are functions of unknown strong interaction phases.
A search for new physics effects in these processes requires a careful
theoretical analysis within the Standard Model of $\Delta S_f$ and $C_f$ 
and not only of $|c'_f/p'_f|$.

Model-independent studies of  the ratios $|c'_f/p'_f|$, providing estimates 
for $\Delta S_f$ for the above final states, were performed in~\cite{CGLRS,GLNQ}. 
Here we will follow Ref.~\cite{GGR,GRZ} in order to obtain  correlated 
bounds directly on $S_f$ and $C_f$ in the two cases of $B^0\to \pi^0K_S$ and 
$B^0\to\eta'K_S$. For simplicity of expressions we will expand the two asymmetries
up to terms linear in $|c'_f/p'_f|$. We will not study theoretical bounds on asymmetries
in $B^0\to\phi K_S$, where one is awaiting a greater consistency between 
BaBar and Belle measurements~\cite{phiK}.

\subsection{$B^0\to\pi^0 K_S$}

Writing
\beq
A(B^0\to \pi^0 K^0) = |p'|e^{i\delta'} - |c'|e^{i\gamma},
\eeq
and denoting $r'\equiv |c'/p'|$, one obtains~\cite{MG} 
\bea
S_{\pi K} & \approx & \sin 2\beta - 2r'\cos 2\beta\sin\gamma\cos\delta',
\nonumber\\
C_{\pi K} & \approx & - 2r'\sin\gamma\sin\delta'.
\label{SCKpi}
\eea
The allowed region in the $(S_{\pi K},
C_{\pi K})$ plane is confined to an ellipse centered at $(\sin 2 \beta, 0)$,
with semi-principal axes $2[r'\sin \gamma]_{\rm max} \cos 2\beta$
and $2[r'\sin\gamma]_{\rm max}$. 
An approximate ellipse providing
these bounds is obtained by relating $B^0\to\pi^0 K^0$ within flavor
SU(3) to $B^0\to\pi^0 \pi^0$ and $B^0\to K^+ K^-$. For simplicity,
we will first neglect the second process given by an exchange-type 
amplitude which is expected to be negligible. It then follows from 
SU(3) that 
\beq
A(B^0\to\pi^0\pi^0) = -\bl|p'|e^{i\delta'} - \bl^{-1}|c'|e^{i\gamma}.
\eeq

Defining a ratio of rates,
\beq
R_{\pi/K} \equiv \frac{\bl^2{\cal B}(B^0 \to \pi^0\pi^0)}
{{\cal B}(B^0 \to \pi^0K^0)},
\eeq
with $R_{\pi/K} = 0.0084 \pm 0.0023$~\cite{HFAG}, 
one has 
\beq\label{RpiK}
R_{\pi/K} = \frac{r'^2 + \bl^4 + 2\bl^2r'\cos\delta'\cos\gamma}
{1 + r'^2 - 2r'\cos\delta'\cos\gamma}.
\eeq 
A scatter plot of $|C_{\pi K}|$ vs. $S_{\pi K}$, 
using (\ref{RpiK}) and exact expressions instead of (\ref{SCKpi}),
is shown in Fig.~4~\cite{GGR}. It describes the allowed region. 
Also shown is the experimental point 
given by a BaBar measurement~\cite{piKBaBar}.
The inner ellipse, which is excluded when neglecting $A(B^0\to K^+K^-)$, 
is still allowed when using the  current upper bound on this amplitude.
In any event, current errors in the asymmetries
are seen to be too large to provide a sensitive test for new physics,
and must be improved for such a test.
\begin{figure}[h]
\includegraphics[height=4.5cm,width=7.0cm]{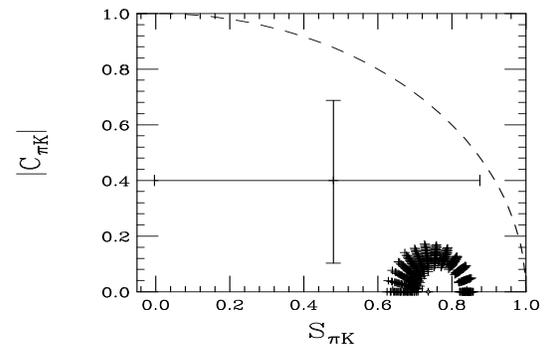}
\caption{Points in $(S_{\pi K},|C_{\pi K}|)$ plane
satisfying (\ref{RpiK}).}
\end{figure}

\subsection{$B^0\to\eta' K_S$}

In order to apply SU(3) to asymmetries in 
$B^0\to\eta'K_S$, where measurements are~\cite{eta'K}, 
\beq
S_{\eta'K}=0.27\pm 0.21,~C_{\eta'K}=0.04\pm 0.13,
\eeq
one may optimize bounds over a whole continuum of combinations of 
corresponding strangeness conserving decays. These include $B^0$ 
decays involving
$\pi^0, \eta$ and $\eta'$ in the final state. Considerable improvements
in some of these branching ratios were achieved recently by 
BaBar~\cite{BaBareta}. The resulting bounds are shown in Fig.~5~\cite{GRZ}.
Regions enclosed by the solid (dashed) curve give current bounds including 
(neglecting) annihilation-type amplitudes, while earlier bounds are given
by the dot-dashed curve. Also shown is a point labeled x denoting the central 
value predicted in a global SU(3) fit to $B\to PP$~\cite{BPP}. 
With recent improvement in bounds on $\Delta S=0$ rates, only a mild 
improvement in the asymmetry measurements is required for achieving 
sensitivity to new physics.
\begin{figure}[h]
\centerline{\includegraphics[height=4.5cm,width=7.0cm]{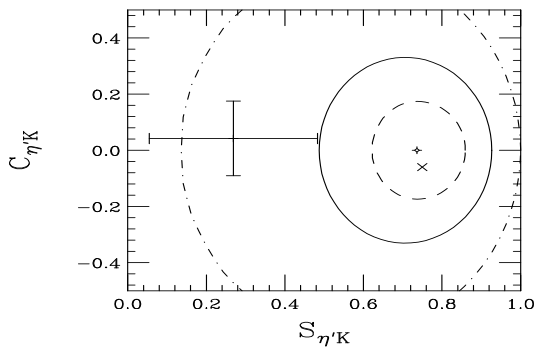}}
\caption{Allowed regions in ($S_{\eta' K},~C_{\eta' K}$) plane.}
\end{figure}

\section{CONCLUSIONS}
 
\begin{itemize}
\item Nothing is as pure and easy as $B\to J/\psi K_S$.
\item The evidence for CP violation in $B\to\pi^+\pi^-$ is consistent with
the KM mechanism,  and is beginning to exclude low values of $\alpha$  
permitted by other CKM bounds.
\item  CP asymmetries in $B^0\to \rho^+\rho^-$ exclude 
values $\alpha < 71^\circ$, in agreement  with CKM constraints. 
Smaller errors in asymmetry measurements may restrict
the range of $\alpha$.
\item Bounds on $\alpha$ from $B\to\rho^{\pm}\pi^{\mp}$ 
overlap nicely with the range obtained from other CKM constraints. 
The intrinsic uncertainty from SU(3) breaking is merely a few degrees.
 \item Current $B^+\to DK^+$ measurements are consistent with CKM 
 constraints on $\gamma$ and require more statistics for stronger bounds.
 \item All this indicates that the KM phase governs 
 CP violation in tree-dominated decays.
 \item Some improvement in asymmetry measurements in 
 $B^0\to \pi^0K_S, \eta'K_S$, and convergence of BaBar and Belle  
 in $B^0\to\phi K_S$, are required for sensitivity to new physics.
\end{itemize}

\medskip
I am grateful to Cheng-Wei Chiang, Jonathan Rosner, Denis Suprun and 
Jure Zupan for enjoyable collaborations.

\def \ajp#1#2#3{Am.\ J. Phys.\ {\bf#1} (#3) #2}
\def \apny#1#2#3{Ann.\ Phys.\ (N.Y.) {\bf#1}, #2 (#3)}
\def \app#1#2#3{Acta Phys.\ Polonica {\bf#1}, #2 (#3)}
\def \arnps#1#2#3{Ann.\ Rev.\ Nucl.\ Part.\ Sci.\ {\bf#1}, #2 (#3)}
\def \art{and references therein}
\def \cmts#1#2#3{Comments on Nucl.\ Part.\ Phys.\ {\bf#1}, #2 (#3)}
\def \cn{Collaboration}
\def \econf#1#2#3{Electronic Conference Proceedings {\bf#1}, #2 (#3)}
\def \efi{Enrico Fermi Institute Report No.}
\def \epjc#1#2#3{Eur.\ Phys.\ J.\ C {\bf#1} (#3) #2}
\def \ib{{\it ibid.}~}
\def \ibj#1#2#3{~{\bf#1}, #2 (#3)}
\def \ijmpa#1#2#3{Int.\ J.\ Mod.\ Phys.\ A {\bf#1} (#3) #2}
\def \ite{{\it et al.}}
\def \jhep#1#2#3{JHEP {\bf#1} (#3) #2}
\def \jpb#1#2#3{J.\ Phys.\ B {\bf#1}, #2 (#3)}
\def \mpla#1#2#3{Mod.\ Phys.\ Lett.\ A {\bf#1} (#3) #2}
\def \nat#1#2#3{Nature {\bf#1}, #2 (#3)}
\def \nc#1#2#3{Nuovo Cim.\ {\bf#1}, #2 (#3)}
\def \nima#1#2#3{Nucl.\ Instr.\ Meth.\ A {\bf#1}, #2 (#3)}
\def \npb#1#2#3{Nucl.\ Phys.\ B~{\bf#1}  (#3) #2}
\def \npps#1#2#3{Nucl.\ Phys.\ Proc.\ Suppl.\ {\bf#1} (#3) #2}
\def \PDG{Particle Data Group, K. Hagiwara \ite, \prd{66}{010001}{2002}}
\def \pisma#1#2#3#4{Pis'ma Zh.\ Eksp.\ Teor.\ Fiz.\ {\bf#1}, #2 (#3) [JETP
Lett.\ {\bf#1}, #4 (#3)]}
\def \pl#1#2#3{Phys.\ Lett.\ {\bf#1}, #2 (#3)}
\def \pla#1#2#3{Phys.\ Lett.\ A {\bf#1}, #2 (#3)}
\def \plb#1#2#3{Phys.\ Lett.\ B {\bf#1} (#3) #2}
\def \prl#1#2#3{Phys.\ Rev.\ Lett.\ {\bf#1} (#3) #2}
\def \prd#1#2#3{Phys.\ Rev.\ D\ {\bf#1} (#3) #2}
\def \prp#1#2#3{Phys.\ Rep.\ {\bf#1} (#3) #2}
\def \ptp#1#2#3{Prog.\ Theor.\ Phys.\ {\bf#1} (#3) #2}
\def \rmp#1#2#3{Rev.\ Mod.\ Phys.\ {\bf#1} (#3) #2}
\def \rp#1{~~~~~\ldots\ldots{\rm rp~}{#1}~~~~~}
\def \yaf#1#2#3#4{Yad.\ Fiz.\ {\bf#1}, #2 (#3) [Sov.\ J.\ Nucl.\ Phys.\
{\bf #1}, #4 (#3)]}
\def \zhetf#1#2#3#4#5#6{Zh.\ Eksp.\ Teor.\ Fiz.\ {\bf #1}, #2 (#3) [Sov.\
Phys.\ - JETP {\bf #4}, #5 (#6)]}
\def \zp#1#2#3{Zeit.\ Phys.\ {\bf#1} (#3) #2}
\def \zpc#1#2#3{Zeit.\ Phys.\ C {\bf#1} (#3) #2 }
\def \zpd#1#2#3{Zeit.\ Phys.\ D {\bf#1}, #2 (#3)}


\begin{thebibliography}{99}

%
\bibitem{psiKs} BaBar \cn, B. Aubert \ite, \prl{89}{201802}{2002};
Belle \cn, K. Abe \ite, \prd{66}{071102}{2002}.

\bibitem{BS}  A. B. Carter and A. I. Sanda, \prd{23}{1567}{1981}; 
I. I. Bigi and A. I. Sanda, \npb{193}{85}{1981}.

\bibitem{KM} M. Kobayashi and T. Maskawa, \ptp{49}{652}{1973}.

\bibitem{MG} M. Gronau, \prl{63}{1451}{1989}.

\bibitem{LonPec} D. London and R. D. Peccei, \plb{223}{257}{1989}; B. Grinstein, 
\plb{229}{280}{1989}.

\bibitem{CKMfitter} J. Charles \ite, hep-ph/0406184.

\bibitem{psiK*} BaBar \cn, reported by M. Verderi at the 39th Rencontres de 
Moriond on Electroweak Interactions and Unified Theories, 21--28 March 2004, 
La Thuile, Aosta, Italy.
    
\bibitem{NP} M. Gronau and D. London, \prd{55}{2845}{1997}; 
Y. Grossman and M. P. Worah, \plb{395}{241}{1997}; D. London 
and A. Soni, \plb{407}{61}{1997}. 

\bibitem{BBNS} M. Beneke, G. Buchalla, M. Neubert, and C. T. Sachrajda,
\npb{606}{245}{2001}; 
C. W. Bauer, D. Pirjol, I. Z. Rothstein and I. W. Stewart, hep-ph/0401188.

\bibitem{tosi} S. Tosi, these proceedings.

\bibitem{schwartz} A. Schwartz, these proceedings.

\bibitem{schrenk} S. Schrenk, these proceedings.

\bibitem{Bapipi} BaBar \cn, presented by H. Jawahery, Proceedings of the XXI 
International Symposium on Lepton and Photon Interactions, Fermilab, Batavia, 
USA, August 11--16 2003.

\bibitem{Bepipi} Belle \cn, K. Abe \ite, \prl{93}{021601}{2004}.

\bibitem{GL} M. Gronau and D. London, \prl{65}{3381}{1990}.

\bibitem{EWP} M. Gronau, D. Pirjol and T. M. Yan, \prd{60}{034021}{1999};
A. Buras and R. Fleischer, \epjc{11}{93}{1999}.

\bibitem{GQ} Y. Grossman and H. R. Quinn, \prd{58}{017504}{1998}; 
J. Charles, \prd{59}{054007}{1999}; M. Gronau, D. London, N. Sinha and 
R. Sinha, \plb{514}{315}{2001}.

\bibitem{HFAG} Heavy Flavor Averaging Group: {\tt http://www.slac.stanford.edu/xorg/hfag}.

\bibitem{Balong} BaBar \cn, B. Aubert \ite, \prd{69}{031102}{2004}.

\bibitem{BarhoSC}  BaBar \cn, B. Aubert \ite, hep-ex/0404029. 

\bibitem{FLNQ} A. F. Falk, Z. Ligeti, Y. Nir and H. R. Quinn, 
\prd{69}{011502(R)}{2004}. 

\bibitem{r0r0} BaBar \cn, B. Aubert \ite, \prl{91}{171802}{2003}. 

\bibitem{Beller+r0} Belle \cn, J. Zhang \ite, \prl{91}{221801}{2003}.

\bibitem{SU3}  D. Zeppenfeld, \zp{8}{77}{1981};
M. Savage and M. Wise, \prd{39}{3346}{1989};
L. L. Chau \ite, \prd{43}{2176}{1991};
B.~Grinstein and R.~F.~Lebed, \prd{53}{6344}{1996}.

\bibitem{GHLR} M.~Gronau, O.~F.~Hernandez, D.~London and J.~L.~Rosner,
\prd{50}{4529}{1994}; \ib {\bf 52} (1995) 6356; \ib {\bf 52} (1995) 6374.

\bibitem{GRpipi} M. Gronau and J. L. Rosner, \plb{595}{339}{2004}.

\bibitem{BPP} C. W. Chiang, M. Gronau, J. L. Rosner and D. A. Suprun, 
hep-ph/0404073.

\bibitem{suprun} D. Suprun, these proceedings.

\bibitem{QCD}  M. Beneke \ite, Ref.~\cite{BBNS}, M. Beneke and M. Neubert, 
\npb{675}{333}{2003}; Y. Y. Keum and A. A. Sanda, \prd{67}{054009}{2003}.

\bibitem{GZ} M. Gronau and J. Zupan, hep-ph/0407002. 

\bibitem{zupan} J. Zupan, these proceedings.

\bibitem{MGPLB} M. Gronau, \plb{233}{479}{1989}. 

\bibitem{BaBarrhopi} BaBar \cn, B.~Aubert {\it et al.}, \prl{91}{201802}{2003}.

\bibitem{CGLRS} C. W. Chiang, M. Gronau, Z. Luo, J. L. Rosner
and D. A. Suprun, \prd{69}{034001}{2004}. 

\bibitem{GW} M. Gronau and D. Wyler, \plb{265}{172}{1991}; M. Gronau 
and D. London, \plb{253}{483}{1991}; M. Gronau, \prd{58}{037301}{1998}.

\bibitem{variants} M. Gronau, Y. Grossman, N. Shuhmaher, A. Soffer and J. 
Zupan, \prd{69}{113003}{2004}.

\bibitem{GGSZ} A. Giri, Y. Grossman, A. Soffer and J. Zupan, \prd{68}
{054018}{2003}.

\bibitem{DKBaBar} BaBar \cn, B. Aubert \ite, hep-ex/0402024.

\bibitem{BDK} M. Gronau, \plb{557}{198}{2003}.

\bibitem{dataDK} Belle \cn, S. K. Swain \ite, \prd{68}{051101}{2003}; 
BaBar \cn, B. Aubert \ite, \prl{92}{202002}{2004}.

\bibitem{BelleDalitz} Belle \cn, A. Poluektov \ite, hep-ex/0406067.

\bibitem{GLNQ} Y. Grossman, Z. Ligeti, Y. Nir and H. R. Quinn,
\prd{68}{015004}{2003}; M. Gronau and J. L. Rosner, \plb{564}{90}{2003};
C. W. Chiang, M. Gronau and J. L. Rosner, \prd{68}{074012}{2003}.

\bibitem{GGR} M. Gronau, Y. Grossman and J.  L. Rosner, 
\plb{579}{331}{2004}.

\bibitem{GRZ} M. Gronau, J. L. Rosner and J. Zupan, \plb{596}{107}{2004}.

\bibitem{phiK} Belle \cn, K. Abe \ite, \prl{91}{261602}{2003};
BaBar \cn, B. Aubert \ite, hep-ex/0403026.

\bibitem{piKBaBar} BaBar \cn, B. Aubert \ite, hep-ph/0403001.

\bibitem{eta'K} BaBar \cn, B. Aubert \ite, \prl{91}{161801}{2003}; 
Belle \cn, K. Abe \ite, \prl{91}{261602}{2003}.

\bibitem{BaBareta} BaBar \cn, B. Aubert \ite, hep-ex/0403025; hep-ex/0403046.


\end{thebibliography}
\end{document}